\pdfoutput=1
\documentclass[12pt, draftclsnofoot, onecolumn]{IEEEtran}
\usepackage[T1]{fontenc}
\usepackage[latin9]{inputenc}
\usepackage{float}
\usepackage{amsmath}
\usepackage{amssymb}
\usepackage{stackrel}
\usepackage{graphicx}
\usepackage{setspace}
\usepackage[unicode=true,
 bookmarks=true,bookmarksnumbered=true,bookmarksopen=true,bookmarksopenlevel=1,
 breaklinks=false,pdfborder={0 0 0},pdfborderstyle={},backref=false,colorlinks=false]
 {hyperref}
\hypersetup{pdftitle={Your Title},
 pdfauthor={Your Name},
 pdfpagelayout=OneColumn, pdfnewwindow=true, pdfstartview=XYZ, plainpages=false}

\makeatletter

\floatstyle{ruled}
\newfloat{algorithm}{tbp}{loa}
\providecommand{\algorithmname}{Algorithm}
\floatname{algorithm}{\protect\algorithmname}

\usepackage[caption=false,font=footnotesize]{subfig}

\makeatother

\begin{document}
\title{An Efficient Ratio Detector for Ambient Backscatter Communication}
\author{Wenjing~Liu,~\IEEEmembership{Graduate Student Member,~IEEE,}~Shanpu~Shen,~\IEEEmembership{Member,~IEEE,}
Danny~H.~K.~Tsang,~\IEEEmembership{Fellow,~IEEE,}~Ranjan~K.~Mallik,~\IEEEmembership{Fellow,~IEEE,}\\
Ross~Murch,~\IEEEmembership{Fellow,~IEEE}\thanks{Manuscript received; This work was supported by the Hong Kong Research
Grants Council under General Research Fund grant 16207620. \textit{(Corresponding
author: Shanpu Shen.)}}\thanks{W. Liu, S. Shen, R. Murch are with the Department of Electronic and
Computer Engineering, The Hong Kong University of Science and Technology,
Clear Water Bay, Kowloon, Hong Kong (e-mail: wliubj@connect.ust.hk;
sshenaa@connect.ust.hk; eermurch@ust.hk).}\thanks{D. H. K. Tsang is with The Hong Kong University of Science and Technology
(Guangzhou), Nansha, Guangzhou, 511400, Guangdong, China, and the
Department of Electronic and Computer Engineering, The Hong Kong University
of Science and Technology, Clear Water Bay, Kowloon, Hong Kong (e-mail:
eetsang@ust.hk).}\thanks{Ranjan K. Mallik is with the Department of Electrical Engineering,
Indian Institute of Technology - Delhi, Hauz Khas, New Delhi 110016,
India (e-mail: rkmallik@ee.iitd.ernet.in).}}
\maketitle
\begin{abstract}
Ambient backscatter communication (AmBC) leverages the existing ambient
radio frequency (RF) environment to implement communication with battery-free
devices. One critical challenge of AmBC systems is signal recovery
because the transmitted information bits are embedded in the ambient
RF signals and these are unknown and uncontrollable. To address this
problem, most existing approaches use averaging-based energy detectors
and consequently the data rate is low and there is an error floor.
Here we propose a new detection strategy based on the ratio between
signals received from a multiple-antenna Reader. The advantage of
using the ratio is that ambient RF signals are removed directly from
the embedded signals without averaging and hence it can increase data
rates and avoid the error floor. Different from original ratio detectors
that use the magnitude ratio of the signals between two Reader antennas,
in our proposed approach, we utilize the complex ratio so that phase
information is preserved and propose an accurate linear channel model
approximation. This allows the application of existing linear detection
techniques from which we can obtain a minimum distance detector and
closed-form expressions for bit error rate (BER). In addition, averaging,
coding and interleaving can also be included to further enhance the
BER. The results are also general, allowing any number of Reader antennas
to be utilized in the approach. Numerical results demonstrate that
the proposed approach performs better than approaches based on energy
detection and original ratio detectors.
\end{abstract}

\begin{IEEEkeywords}
Ambient backscatter communication, averaging, channel linearization,
ratio detector, repetition code
\end{IEEEkeywords}

\section{Introduction}

\IEEEPARstart{T}{he} Internet of things (IoT) has attracted significant
attention in both academia and industrial circles \cite{iot_survey_al}.
Its function is to provide ubiquitous connectivity among a large number
of small IoT computing devices (associated with people, homes, vehicles,
and other factors) embedded in the environment and on and within people
\cite{backscatter_survery_REzaei}. Powering the IoT devices is however
challenging. Batteries have a limited lifetime and require significant
maintenance costs to replace, while the use of solar energy requires
the devices to be visible and in lighted regions.

One approach to meet the challenge of powering IoT devices is ambient
backscatter communication (AmBC) which was first proposed by Liu \textit{et.
al} in 2013 \cite{thin_Liu_2013}. In an AmBC system, a ``Tag''
harvests energy from ambient radio frequency (RF) signals to power
its circuits \cite{ShanpuShen2017_TAP_EHPIXEL}, \cite{directional_shanpu_2021}.
It transmits its data to a ``Reader'' by tuning its antenna impedance
to reflect the received RF signals. Specifically, the Tag maps ``0''
and ``1'' bits to RF waveforms by adjusting the load impedance of
the antenna between absorbing and reflecting states \cite{survery_VanHuynh_2018}.
Different from conventional backscatter communication, e.g., Radio
Frequency Identification (RFID), where a dedicated source of RF radiation
is needed, AmBC enables battery-free Tags to communicate with other
devices by harnessing ambient RF signals emitted from existing wireless
systems (such as DTV \cite{thin_Liu_2013}, FM \cite{FMbackscatter_Wang_2017},
\cite{AmBC_FM_Daskalakis_2017}, and Wi-Fi \cite{WiFiback_Kwllogg_2014}-\nocite{Backfi_Bharadia_2015}\nocite{passiveWIFI_Kellogg_2016}\cite{Efficient_BackFi_Ji_2019}).
Thus, AmBC can operate without any dedicated RF source or extra frequency
spectrum allocation, making it a promising approach for realizing
a sustainable IoT ecosystem \cite{backscatter_survery_REzaei}, \cite{survery_VanHuynh_2018}.
In addition, the Reader does not require special duplexing circuitry
(as in RFID Readers) so that existing devices, such as Wi-Fi, can
potentially be modified to be Readers of AmBC signals.

In AmBC systems, the Reader receives two types of signals: the direct
link signal emitted from the ambient RF source and the signal backscattered
by the Tag. The backscattered signal is much weaker than the direct
link signal due to the round-trip loss \cite{Constellation_Learning_Based_Zhang_2019}.
In addition, the ambient RF signal is unknown and uncontrollable.
Therefore, the detection of AmBC signals transmitted by the Tag is
challenging.

To tackle this challenge, averaging-based energy detection is a widely
used method to recover the Tag data. The basic principle of the energy
detector is to average the power of the received signal and obtain
its power level, which can be thresholded for detecting Tag symbols.
In \cite{thin_Liu_2013}, the feasibility of communication between
two battery-free devices utilizing ambient TV signals has been shown
using averaging-based energy detection. The bit error rate (BER) performance
of the energy detector with differential encoding and maximum-likelihood
(ML) detector has also been investigated \cite{Detection_perfprmance_Wang_2016}-\nocite{uplink_wang_2017}\cite{Semi_coherent_Qian_2017}.
In \cite{Nocoherent_Qian_2017}, a joint energy detection scheme has
been adopted where the differential coding characteristics have been
exploited to avoid channel estimation and pilot symbols. The non-coherent
energy detector has also been generalized to multiple-phase shift
keying (MPSK) \cite{MPSK_Qian_2019}. In \cite{Manchester_coding_Tao_2018},
Manchester coding has been applied and an optimal non-coherent detector
has been analyzed \cite{noncoherent_tvt_2020}.

There are several disadvantages with the averaging-based energy detection
approach. The first is that the averaging process significantly reduces
the resulting bit rate. The averaging-based energy detection also
has a severe error floor problem due to direct link interference \cite{optimal_multiple_reader_Tao_2020}.
That is, the BER converges to a nonzero floor with increase in the
transmitted power from the ambient RF source \cite{multiple_antenna_cognitive_Guo_2019}.
Furthermore, energy detection generally suffers from performance loss
since the averaging operation loses all the phase information \cite{Constellation_Learning_Based_Zhang_2019}.

Due to the limitations of averaging-based energy detection, enhancements
to AmBC detection techniques have also been investigated. This includes
implementing multiple antennas at the Reader to provide a straightforward
method that can efficiently improve link quality and reliability and
significantly increase the data rate. In \cite{multiple_energy_ma_2015},
\cite{Nocohrent_OFDM_ElMossallamy_2019}, blind energy detectors with
multi-antenna Readers have been proposed. While the BER is lowered
by increasing the number of antennas, the error floor caused by energy
detectors still exists. Alternative approaches have also been considered,
and this includes isolating the spectrum of the backscattered and
ambient RF signals \cite{MFSK_Tao_2019}, \cite{Swiching_frequency_Vougioukas_2019},
statistical clustering \cite{multiple_antenna_cognitive_Guo_2019},
and statistical covariances \cite{covariance_zeng_2016}.

A very different approach to remove the ambient RF background signal
from the transmitted bits has also been suggested in 2014, which is
based on using the ratio of signals from separate antennas at the
Reader \cite{umo_Parks_2014}. It has been demonstrated that taking
the ratio of two antenna branches at the Reader can cancel the ambient
RF source, allowing increases in data rate \cite{umo_Parks_2014}.
Ratio detection is particularly suited to AmBC configurations that
leverage systems such as Wi-Fi and where the AmBC Reader is integrated
with the Wi-Fi station. This is because the ratio detector for AmBC
can straightforwardly leverage the baseband outputs and synchronization
from the Wi-Fi system to form the required ratio. The idea has been
further investigated by finding an approximate detection threshold
and the corresponding closed-form BER expression has been derived
\cite{expand_umo_Ma_2018}. As there is no error floor and data rates
can be increased, the ratio detector approach has significant potential
for further development.

In this paper, we take a new look at the AmBC Reader with ratio detection
to overcome the drawbacks of conventional averaging-based energy detection.
Our new contributions include:

1) Proposing a new ratio detector: Different from original ratio detectors
\cite{umo_Parks_2014}, \cite{expand_umo_Ma_2018} that use magnitude
information only, we propose using the direct ratio of the received
signals so that the phase information is preserved. The probability
density functions (PDFs) of the resulting statistic and optimal ML
detector are also derived analytically.

2) Performing channel linearization: Using the proposed ratio allows
us to develop an accurate channel linearization for the AmBC system.
Based on the linear channel model, we simplify the ML detector to
a minimum distance detector. The closed-form BER expression of the
detector based on the proposed channel model is also presented.

3) Proposing averaging and coding techniques: Another advantage of
having an accurate linear channel model for the AmBC system is that
it provides a straightforward approach to averaging and coding. We
show that the use of averaging, coding, and interleaving can further
enhance the detection performance. Coding with and without interleaving
is shown to be more effective than averaging as expected.

4) Generalizing to more than two antenna systems: The use of the ratio
detector opens up the use of many types of ratios and their combinations
when more than two antennas are utilized. We describe a straightforward
ratio selection approach that efficiently lowers the BER for more
than two antennas.

5) Numerical results: Comparisons between the proposed ratio detector,
the conventional averaging-based energy detector, and the original
ratio detector show the enhanced performance of our technique. For
example, when the direct link signal-to-noise ratio (SNR) is $15$
dB and $M=1000$, the BER of the proposed detector is almost 100 times
lower than averaging-based energy detection. The effectiveness of
the proposed ratio selection scheme for more than two antennas is
also shown.

In Section II, the system model of the proposed AmBC system with a
multiple-antenna Reader is provided. In Section III, we introduce
our proposed ratio detector. The resulting accurate linear channel
approximation, minimum distance detector, and exact closed-form BER
expressions of the detector are detailed. Section IV introduces averaging,
coding, and interleaving while Section V generalizes the approach
to arbitrary numbers of antennas. Section VI provides simulation results
and the conclusion is given in Section VII.

\textit{Notation:} Bold lower and upper case letters denote vectors
and matrices, respectively. A symbol not in bold font represents a
scalar. $x^{*}$, $\mathfrak{R}\left\{ x\right\} $, and $\left|x\right|$
refer to the conjugate, real part, and modulus, respectively, of a
complex scalar $x$. $\mathbf{x}^{T}$ refers to the transpose of
a vector $\mathbf{x}$. $\mathrm{diag}\left(\mathbf{x}\right)$, returns
a square diagonal matrix with the elements of vector \textbf{$\mathbf{x}$}
on the main diagonal. $\mathbb{R}^{M}$ denotes the space of $M$-dimensional
real vectors. $\mathbf{X}^{T}$ refers to the transpose of a matrix
$\mathbf{X}$. $\mathcal{CN}\left(\mu,\sigma^{2}\right)$ refers to
the circularly symmetric complex Gaussian distribution with mean $\mu$
and variance $\sigma^{2}$. $f\left(\cdot\right)$ denotes the PDF
of a random variable and $\Pr\left(\cdot\right)$ denotes the probability
of an event.

\section{System Model}

Consider an AmBC system consisting of an ambient RF source, a passive
Tag equipped with a single antenna, and a Reader equipped with $Q$
antennas as shown in Fig. 1. The ambient RF signal is denoted by $s\left(n\right)$
(with symbol period $T_{a}$), where $n=1,2,\cdots$ is the symbol
index of the ambient source. $s\left(n\right)$ is assumed random
(and may come from different RF sources) and follows a circularly
symmetric complex Gaussian distribution $\mathcal{CN}\left(0,P_{s}\right)$,
where $P_{s}$ denotes the average power of the ambient RF signal.

For the $n$th symbol period of the ambient source, the total signal
received by the $q$th Reader antenna can be expressed as
\begin{align}
z_{q}\left(n\right) & =\left(A_{\mathrm{SR}}h_{q}^{\mathrm{SR}}+\alpha A_{\mathrm{TR}}A_{\mathrm{ST}}h_{q}^{\mathrm{TR}}h^{\mathrm{ST}}x\right)s\left(n\right)+w_{q}\left(n\right),\label{eq:zq(n)}
\end{align}
where $A_{\mathrm{TR}}$ and $h_{q}^{\mathrm{TR}}$ denote the large-
and small-scale channel fading between the Tag antenna and the $q$th
Reader antenna, $A_{\mathrm{ST}}$ and $h^{\mathrm{ST}}$ denote the
large- and small-scale channel fading between the ambient RF source
and the Tag antenna, $A_{\mathrm{SR}}$ and $h_{q}^{\mathrm{SR}}$,
respectively, denote the large- and small-scale channel fading between
the ambient RF source and the $q$th Reader antenna. The symbol transmitted
by the Tag antenna is denoted as $x$, and $w_{q}\left(n\right)$
is the additive white Gaussian noise (AWGN) at the $q$th Reader antenna.
Besides, we assume block-fading, where the channel coherence time
exceeds the transmission duration of a block. However, the channel
will still vary from one block to another because multi-path components
are varying over time \cite{coherent_bpsk_wang}.

\begin{figure}[t]
\centering{}\includegraphics[scale=0.38]{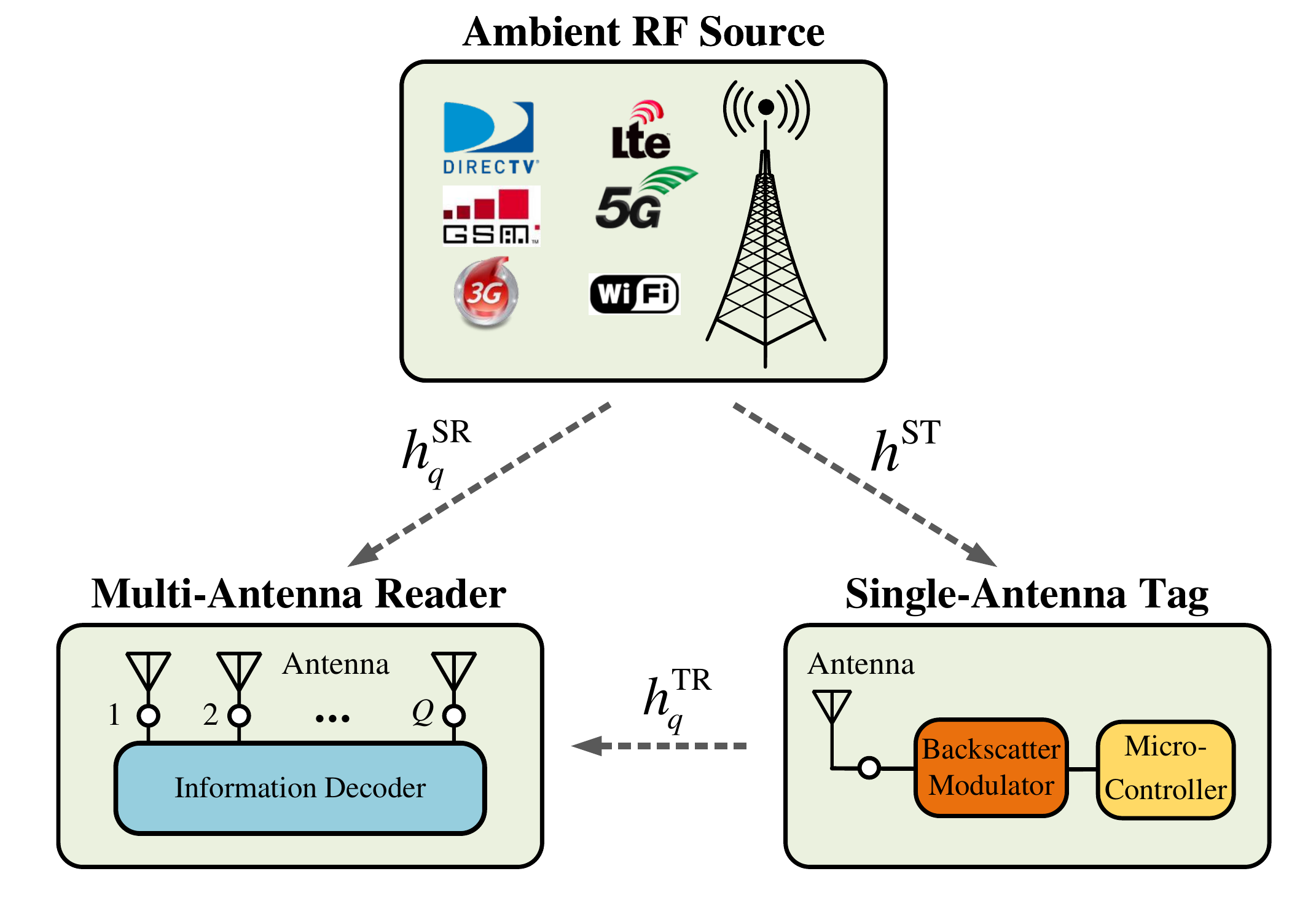}\caption{System model of an AmBC system with a single-antenna Tag and a multiple-antenna
Reader.}
\end{figure}

We assume $h_{q}^{\mathrm{SR}}$, $h_{q}^{\mathrm{TR}}$ $\forall$
$q$, and $h^{\mathrm{ST}}$ are independent and identically distributed
(i.i.d.), quasi-static, and frequency flat. We also assume $A_{\mathrm{SR}}\approx A_{\mathrm{ST}}$
since the Tag and the Reader are close to each other. Therefore, we
can normalize \eqref{eq:zq(n)} by $A_{\mathrm{SR}}$ to equivalently
rewrite \eqref{eq:zq(n)} as
\begin{align}
\bar{z}_{q}\left(n\right) & =\left(h_{q}^{\mathrm{SR}}+h_{q}^{\mathrm{TR}}gx\right)s\left(n\right)+\bar{w}_{q}\left(n\right),\label{eq:normalized Zq(n)}
\end{align}
where we define $\bar{z}_{q}\left(n\right)=z{}_{q}\left(n\right)/A_{\mathrm{SR}}$,
$g=\alpha A_{\mathrm{TR}}h^{\mathrm{ST}}$ and $\bar{w}_{q}\left(n\right)=w_{q}\left(n\right)/A_{\mathrm{SR}}$.
We assume $\bar{w}_{q}\left(n\right)\sim\mathcal{CN}\left(0,N_{w}\right)$
where $N_{w}$ denotes the normalized noise power.

Leveraging \eqref{eq:normalized Zq(n)}, we can define the direct
link SNR $\gamma_{d}$ as
\begin{equation}
\gamma_{d}\triangleq\frac{P_{s}}{N_{w}},
\end{equation}
which is the ratio of ambient RF signal power and normalized noise
power at the Reader. This is referred to as SNR in the remainder of
the paper. We also define a relative SNR $\Delta\gamma$ as
\begin{equation}
\Delta\gamma\triangleq\frac{1}{\alpha^{2}A_{\mathrm{TR}}^{2}},
\end{equation}
which is the ratio of the signal power from the direct link and the
backscatter link. The relative SNR $\Delta\gamma$ is usually greater
than 30 dB and therefore special techniques need to be utilized for
AmBC signal detection \cite{Wenjing_FIRST_2021}.

The most common detection approach in AmBC is averaging-based energy
detection but it suffers from performance issues as discussed earlier.
As an alternative approach, it is possible to consider the ratio between
two antenna branches (say $q=1$ and 2) as it can effectively cancel
or divide out the unknown ambient RF source signal $s\left(n\right)$.
Not only can this overcome the error floor of conventional averaging-based
energy detection, it can also increase the data rate. The originally
proposed ratio detector uses the magnitude ratio of the received signals
at two antenna branches \cite{umo_Parks_2014,expand_umo_Ma_2018}.
It is shown that the ratio detector avoids the error floor phenomenon
faced by averaging-based energy detector \cite{expand_umo_Ma_2018},
with the data rate and communication range having been greatly improved
\cite{umo_Parks_2014}. However, taking the magnitude ratio loses
potentially useful phase information, and also makes the channel model
nonlinear. In \cite{umo_Parks_2014}, it is pointed out that the use
of magnitudes allows the use of non-coherent receivers. However, with
AmBC Readers being integrated in systems such as Wi-Fi, signals received
by different antenna branches will be phase coherent due to the use
of a single local oscillator. This allows the consideration of complex
signal ratios. An advantage of using complex ratios is that an accurate
linear channel model can be developed for AmBC as shown next.

\section{Proposed Ratio Detector}

We firstly analyze the system performance of the ratio detector with
$Q=2$ antenna branches to highlight its functionality. Later in Section
V we provide a straightforward generalization for $Q>2$ antennas.

We denote the signal ratio between the two antenna $(q=1,2)$ branches
as
\begin{equation}
\lambda\left(n\right)=\frac{\bar{z}_{1}\left(n\right)}{\bar{z}_{2}\left(n\right)}=\frac{\left(h_{1}^{\mathrm{SR}}+h_{1}^{\mathrm{TR}}gx\right)s\left(n\right)+\bar{w}_{1}\left(n\right)}{\left(h_{2}^{\mathrm{SR}}+h_{2}^{\mathrm{TR}}gx\right)s\left(n\right)+\bar{w}_{2}\left(n\right)}.\label{eq:advanced_channel_model}
\end{equation}
This is different from the originally proposed ratio detector as we
have removed the magnitude or modulus operation. The advantage of
using this ratio is that it allows us to develop an accurate linear
channel model for the AmBC system. Subsequently, the use of linear
detectors, averaging, and coding can be utilized.

One issue in forming the ratio \eqref{eq:advanced_channel_model}
is retaining its phase. However, proposed AmBC Readers are likely
to be integrated with existing wireless systems such as Wi-Fi. In
these circumstances, the system will be coherent and phase coherence
will exist between antenna branches. Therefore, extracting the I and
Q signals for each antenna branch and taking the ratio will not likely
require additional information. This is a significant advantage of
using this ratio.

\subsection{PDF and ML detection}

In this paper, for simplicity we focus on using BPSK modulation for
the transmit signal $x$, that is, $x=\pm1$. Other common modulation
techniques such as ASK, PSK and QAM, can also be applied in the proposed
ratio detector. Defining $\mu_{q}=h_{q}^{\mathrm{SR}}+h_{q}^{\mathrm{TR}}gx$,
$q=1,2,$ as the composite channel between the Reader and Tag when
$x=\pm1$ allows us to write $\mu_{q}s\left(n\right)+\bar{w}_{q}\left(n\right)\sim\mathcal{CN}\left(0,\sigma_{q}^{2}\right)$
where $\sigma_{q}^{2}=\left|\mu_{q}\right|^{2}P_{s}+N_{w}$, $q=1,2$.

\begin{figure}[t]
\begin{centering}
\includegraphics[scale=0.48]{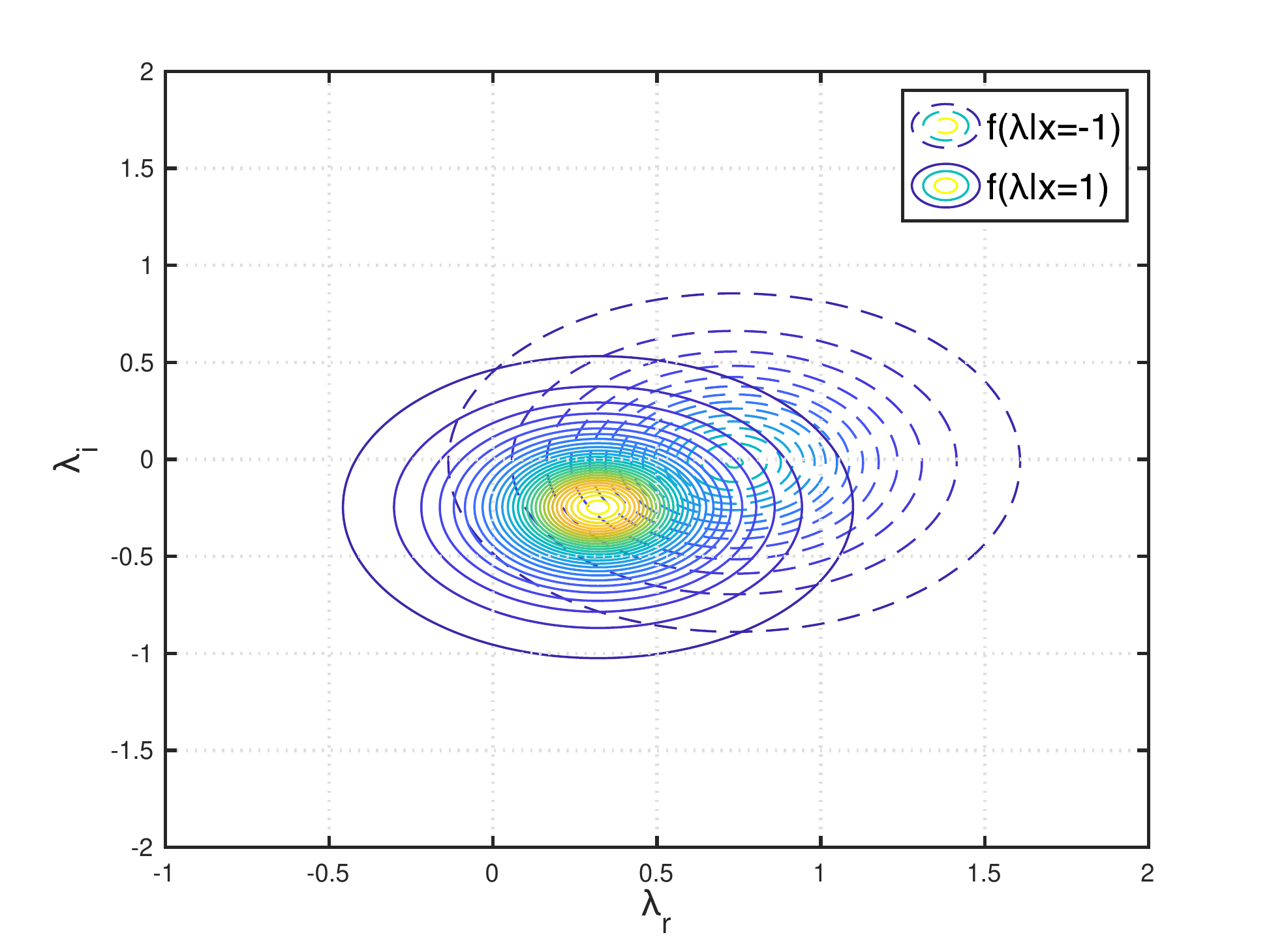}
\par\end{centering}
\centering{}\caption{\label{fig:PDF_new_ratio}Example PDFs of two backscatter states for
the proposed ratio detector when $\gamma_{d}=10$ dB.}
\end{figure}

Since $\lambda\left(n\right)$ is a ratio of two complex Gaussian
distributed random variables, according to \cite{Baxley2010_ratioCN_zeromean,yan2018_ratioCN_generalized},
its PDF is given by
\begin{equation}
f\left(\lambda\right)=\frac{1-\left|\rho\right|^{2}}{\pi\sigma_{1}^{2}\sigma_{2}^{2}}\left(\frac{\left|\lambda\right|^{2}}{\sigma_{1}^{2}}+\frac{1}{\sigma_{2}^{2}}-2\frac{\rho_{r}\lambda_{r}-\rho_{i}\lambda_{i}}{\sigma_{1}\sigma_{2}}\right)^{-2},\label{eq:f_lambda}
\end{equation}
where $\lambda_{r}$ and $\lambda_{i}$ represent the real and imaginary
parts of $\lambda$, respectively; $\rho_{r}$ and $\rho_{i}$ represent
the real and imaginary parts, respectively, of $\rho$, which is the
complex correlation coefficients between $\bar{z}_{1}\left(n\right)$
and $\bar{z}_{2}\left(n\right)$ and is written as
\begin{equation}
\rho=\frac{\mu_{1}^{*}\mu_{2}P_{s}}{\sigma_{1}\sigma_{2}}.
\end{equation}
 The resulting PDFs for an example channel realization with $f\left(\lambda\left|x=-1\right.\right)$
and $f\left(\lambda\left|x=1\right.\right)$ are shown in Fig. \ref{fig:PDF_new_ratio}
when the direct link SNR is $10$ dB.

In ML detection, the goal is to design an optimal detector that can
minimize the error probability, or equivalently, maximize the correct
decision probability
\begin{align}
\hat{x} & =\underset{x\in\left\{ -1,1\right\} }{\arg\max}\:\Pr\left(x\left|\lambda\right.\right).\label{eq:map}
\end{align}
Since the transmit messages $x=-1$ and $x=1$ are equiprobable, according
to Baye's theorem, the ML criterion is written as
\begin{equation}
\hat{x}=\begin{cases}
-1, & \textrm{ if }f\left(\lambda\left|x=-1\right.\right)\geqslant f\left(\lambda\left|x=1\right.\right)\\
1, & \textrm{ if }f\left(\lambda\left|x=-1\right.\right)<f\left(\lambda\left|x=1\right.\right)
\end{cases}.\label{eq:ML_before_linearize}
\end{equation}
That is, we substitute the obtained $\lambda$ in \eqref{eq:f_lambda},
and then determine $x$ by comparing the size relationship of two
posterior density functions. Channel state information (CSI) is also
required for evaluating \eqref{eq:ML_before_linearize}. Several channel
estimation methods have been demonstrated previously that can provide
accurate CSI \cite{Blind_channel_estimation_Ma_2018}-\nocite{Machine_CSI_mA_2018,Sparse_CSI_kIM_2018,CSI_massive_Reader_Zhao_2019,joint_CSI_Darsena_2018}\cite{deep_csi_ambc_xuemeng}
and in the remainder of this paper we assume the Reader has perfect
CSI.

The optimal ML detector is computationally intricate since it needs
to compute intricate PDFs and it is not linear. It can also be observed
that determining a signal threshold for detection is difficult. The
intricacy of ML detection can be completely avoided by utilizing a
different approach. It is shown next that the channel model \eqref{eq:advanced_channel_model}
can be accurately approximated by a linear channel model under the
conditions of AmBC. This allows us to devise a straightforward minimum
distance detector. In addition it opens up the use of all of the linear
detection tools available such as averaging, coding, and selection
diversity.

\subsection{Linearization of Proposed Ratio Detector}

Utilizing the complex ratio \eqref{eq:advanced_channel_model}, we
develop an accurate channel linearization for AmBC. We rewrite the
ratio $\bar{z}_{1}\left(n\right)\left/\bar{z}_{2}\left(n\right)\right.$
as
\begin{align}
\frac{\bar{z}_{1}\left(n\right)}{\bar{z}_{2}\left(n\right)} & =\frac{h_{1}^{\mathrm{SR}}\left(1+\frac{h_{1}^{\mathrm{TR}}gx}{h_{1}^{\mathrm{SR}}}+\frac{\bar{w}_{1}\left(n\right)}{h_{1}^{\mathrm{SR}}s\left(n\right)}\right)}{h_{2}^{\mathrm{SR}}\left(1+\frac{h_{2}^{\mathrm{TR}}gx}{h_{2}^{\mathrm{SR}}}+\frac{\bar{w}_{2}\left(n\right)}{h_{2}^{\mathrm{SR}}s\left(n\right)}\right)}.\label{eq:ratio_without_modular}
\end{align}
Taking logarithms we have
\begin{align}
\log\left(\frac{\bar{z}_{1}\left(n\right)}{\bar{z}_{2}\left(n\right)}\right) & =\log\left(\frac{h_{1}^{\mathrm{SR}}}{h_{2}^{\mathrm{SR}}}\right)+\log\left(1+\frac{h_{1}^{\mathrm{TR}}gx}{h_{1}^{\mathrm{SR}}}+\frac{\bar{w}_{1}\left(n\right)}{h_{1}^{\mathrm{SR}}s\left(n\right)}\right)-\log\left(1+\frac{h_{2}^{\mathrm{TR}}gx}{h_{2}^{\mathrm{SR}}}+\frac{\bar{w}_{2}\left(n\right)}{h_{2}^{\mathrm{SR}}s\left(n\right)}\right).\label{eq:logarithm}
\end{align}

Under typical AmBC conditions, the direct link (ambient RF source
to Reader) has a significantly higher channel gain than the indirect
link (ambient RF source to Tag and then backscattered to Reader).
This corresponds to the fact that the relative SNR $\Delta\gamma$
in practical AmBC systems is usually higher than 30 dB \cite{Wenjing_FIRST_2021}.
Therefore, with $q=1,2$, the signals very frequently satisfy the
conditions
\begin{equation}
\left|h_{q}^{\mathrm{TR}}gx\right|\ll\left|h_{q}^{\mathrm{SR}}\right|;\label{eq:channel property 1}
\end{equation}
\begin{equation}
\left|\bar{w}_{q}\left(n\right)\right|\ll\left|s\left(n\right)h_{q}^{\mathrm{SR}}\right|.\label{eq:channel property 2}
\end{equation}
Under these conditions it can be readily seen that
\begin{equation}
\left|\frac{h_{q}^{\mathrm{TR}}gx}{h_{q}^{\mathrm{SR}}}+\frac{\bar{w}_{q}\left(n\right)}{h_{q}^{\mathrm{SR}}s\left(n\right)}\right|\rightarrow0.\label{eq:channel property 3}
\end{equation}
Using the limit that $\log\left(1+x\right)\rightarrow x$ when $x\rightarrow0$
and the property \eqref{eq:channel property 3}, we have
\begin{equation}
\log\left(1+\frac{h_{q}^{\mathrm{TR}}gx}{h_{q}^{\mathrm{SR}}}+\frac{\bar{w}_{q}\left(n\right)}{h_{q}^{\mathrm{SR}}s\left(n\right)}\right)\rightarrow\frac{h_{q}^{\mathrm{TR}}gx}{h_{q}^{\mathrm{SR}}}+\frac{\bar{w}_{q}\left(n\right)}{h_{q}^{\mathrm{SR}}s\left(n\right)}.
\end{equation}
Thus, \eqref{eq:logarithm} can be expressed as
\begin{align}
\log\left(\frac{\bar{z}_{1}\left(n\right)}{\bar{z}_{2}\left(n\right)}\right) & =\log\left(\frac{h_{1}^{\mathrm{SR}}}{h_{2}^{\mathrm{SR}}}\right)+\left(\frac{h_{1}^{\mathrm{TR}}}{h_{1}^{\mathrm{SR}}}-\frac{h_{2}^{\mathrm{TR}}}{h_{2}^{\mathrm{SR}}}\right)gx+\left(\frac{\bar{w}_{1}\left(n\right)}{h_{1}^{\mathrm{SR}}}-\frac{\bar{w}_{2}\left(n\right)}{h_{2}^{\mathrm{SR}}}\right)\frac{1}{s\left(n\right)}.\label{eq:linearize_with_bias}
\end{align}

The constant bias term $\log\left(h_{1}^{\mathrm{SR}}\left/h_{2}^{\mathrm{SR}}\right.\right)$
in \eqref{eq:linearize_with_bias} can be estimated and removed. Let
us define $\hat{y}\triangleq\log\left(\bar{z}_{1}\left(n\right)\left/\bar{z}_{2}\left(n\right)\right.\right)-\log\left(h_{1}^{\mathrm{SR}}\left/h_{2}^{\mathrm{SR}}\right.\right)$;
according to \eqref{eq:linearize_with_bias}, we can obtain
\begin{equation}
\hat{y}=hx+\tilde{w},\label{eq:linearized model_without_phase}
\end{equation}
where $h$ is defined as
\begin{equation}
h\triangleq\left(\frac{h_{1}^{\mathrm{TR}}}{h_{1}^{\mathrm{SR}}}-\frac{h_{2}^{\mathrm{TR}}}{h_{2}^{\mathrm{SR}}}\right)g,\label{eq:h}
\end{equation}
and $\tilde{w}$ is defined as
\begin{equation}
\tilde{w}\triangleq\left(\frac{\bar{w}_{1}\left(n\right)}{h_{1}^{\mathrm{SR}}}-\frac{\bar{w}_{2}\left(n\right)}{h_{2}^{\mathrm{SR}}}\right)\frac{1}{s\left(n\right)}.\label{eq:w}
\end{equation}
Note that $\left(\bar{w}_{1}\left(n\right)\left/h_{1}^{\mathrm{SR}}\right.-\bar{w}_{2}\left(n\right)\left/h_{2}^{\mathrm{SR}}\right.\right)$
still follows a complex Gaussian distribution, so $\tilde{w}$ is
a ratio of two complex Gaussian variables. According to \cite{Baxley2010_ratioCN_zeromean},
the PDF of $\tilde{w}$ can be obtained as
\begin{align}
f\left(\tilde{w}\right) & =\frac{N_{w}}{\pi P_{s}}\left(\frac{1}{\left|h_{1}^{\mathrm{SR}}\right|^{2}}+\frac{1}{\left|h_{2}^{\mathrm{SR}}\right|^{2}}\right)\left(\left|\tilde{w}\right|^{2}+\left(\frac{1}{\left|h_{1}^{\mathrm{SR}}\right|^{2}}+\frac{1}{\left|h_{2}^{\mathrm{SR}}\right|^{2}}\right)\frac{N_{w}}{P_{s}}\right)^{-2}.
\end{align}

A further detail is that herein we only take the principal value of
each logarithm. As a result, the phase of $\log\left(\bar{z}_{1}\left(n\right)\left/\bar{z}_{2}\left(n\right)\right.\right)$
is restricted to $\left(-\pi,\left.\pi\right]\right.$ while the right-hand
side of \eqref{eq:logarithm} has a phase in $\left(-3\pi,\left.3\pi\right]\right.$.
Therefore, it is possible that the phase of the right-hand side of
\eqref{eq:logarithm} has $\pm2\pi$ shift with $\log\left(\bar{z}_{1}\left(n\right)\left/\bar{z}_{2}\left(n\right)\right.\right)$.
Consequently, \eqref{eq:linearized model_without_phase} does not
always hold. Thus, we cannot utilize \eqref{eq:linearized model_without_phase}
directly to detect the Tag data since the phase shift will decrease
the detection performance. In order to reduce its influence, we propose
a phase compensation method.

Recall that the direct link has a significantly higher channel gain
than the indirect link. Thus $\log\left(h_{1}^{\mathrm{SR}}\left/h_{2}^{\mathrm{SR}}\right.\right)$
is dominant in $\log\left(\bar{z}_{1}\left(n\right)\left/\bar{z}_{2}\left(n\right)\right.\right)$
and we can compare the phase of them to reveal if there is a $2\pi$
phase shift. After removing the bias $\log\left(h_{1}^{\mathrm{SR}}\left/h_{2}^{\mathrm{SR}}\right.\right)$,
the extra phase $\Delta\phi$ that needs to be compensated on $\hat{y}$
is
\begin{equation}
\Delta\phi=\begin{cases}
2\pi & ,\textrm{if}\:\phi\left(\log\left(\frac{\bar{z}_{1}\left(n\right)}{\bar{z}_{2}\left(n\right)}\right)\right)-\phi\left(\log\left(\frac{h_{1}^{\mathrm{SR}}}{h_{2}^{\mathrm{SR}}}\right)\right)<-\pi\\
-2\pi & ,\textrm{if}\:\phi\left(\log\left(\frac{\bar{z}_{1}\left(n\right)}{\bar{z}_{2}\left(n\right)}\right)\right)-\phi\left(\log\left(\frac{h_{1}^{\mathrm{SR}}}{h_{2}^{\mathrm{SR}}}\right)\right)>\pi\\
0 & ,\textrm{\ensuremath{\mathrm{otherwise}}}
\end{cases},\label{eq:phase shift}
\end{equation}
where $\phi\left(\log\left(\bar{z}_{1}\left(n\right)\left/\bar{z}_{2}\left(n\right)\right.\right)\right)$
and $\phi\left(\log\left(h_{1}^{\mathrm{SR}}\left/h_{2}^{\mathrm{SR}}\right.\right)\right)$
represent the phase of $\log\left(\bar{z}_{1}\left(n\right)\left/\bar{z}_{2}\left(n\right)\right.\right)$
and $\log\left(h_{1}^{\mathrm{SR}}\left/h_{2}^{\mathrm{SR}}\right.\right)$,
respectively.

The modified \eqref{eq:linearized model_without_phase} is then written
as
\begin{equation}
y=\hat{y}+\jmath\Delta\phi=hx+\tilde{w},\label{eq:linear channel model}
\end{equation}
which is the finalized linear channel model for the proposed ratio
detector and $\jmath=\sqrt{-1}$. The effectiveness of phase compensation
will be shown in Section VI.

In summary, by performing the logarithm approximation \eqref{eq:linearize_with_bias},
removing bias, and compensating phase shift \eqref{eq:phase shift},
we can approximate the ratio channel model in AmBC \eqref{eq:advanced_channel_model}
as a linear channel model \eqref{eq:linear channel model}. This channel
model approximation is accurate for most practical configurations
in AmBC as shown later in Section VI.

\subsection{Minimum Distance Detector}

The computational complexity of the optimal ML detector \eqref{eq:ML_before_linearize}
can be reduced by leveraging the linearized AmBC channel model to
simplify the detection process.

For simplicity, we further define $\tau$ as
\begin{equation}
\tau\triangleq\left(\frac{1}{\left|h_{1}^{\mathrm{SR}}\right|^{2}}+\frac{1}{\left|h_{2}^{\mathrm{SR}}\right|^{2}}\right)\frac{N_{w}}{\pi P_{s}}.\label{eq:tao}
\end{equation}
The ML detector is then given by
\begin{align}
\hat{x} & =\underset{x\in\left\{ -1,1\right\} }{\arg\max}\:f\left(y\left|x\right.\right)\nonumber \\
 & =\underset{x\in\left\{ -1,1\right\} }{\arg\max}\:\tau\left(\left|y-hx\right|^{2}+\pi\tau\right)^{-2}.\label{f(y|xi)}
\end{align}
It can be readily checked that the function to be maximized in \eqref{f(y|xi)}
is monotonically decreasing with $\left|y-hx\right|^{2}$. Thus \eqref{f(y|xi)}
can be simplified as
\begin{equation}
\hat{x}=\underset{x\in\left\{ -1,1\right\} }{\arg\min}\:\left|y-hx\right|^{2},\label{eq:minimum distance}
\end{equation}
where $h$ is given by \eqref{eq:h}. It implies that the ML detector
\eqref{eq:ML_before_linearize} can be reduced to a minimum distance
detector.

To conclude, based on the proposed linearized AmBC channel model \eqref{eq:linear channel model},
we can apply linear detection to detect $x$ and avoid computing the
complicated PDF functions \eqref{eq:f_lambda}. Therefore, the computational
complexity for detection is greatly reduced without a reduction in
performance as shown in Section VI.

\subsection{BER Analysis}

Utilizing the linear AmBC channel model, we can provide a BER analysis
for the proposed minimum distance detector.

Suppose we transmit $x$ in a single transmission. According to \eqref{eq:minimum distance},
the error event is the distance between the received signal $y$ and
$hx$ is larger than $-hx$. Equivalently,
\begin{equation}
\left|y-hx\right|^{2}>\left|y+hx\right|^{2};
\end{equation}
in other words,
\begin{equation}
\mathfrak{R}\left\{ hx\tilde{w}\right\} <-\left|h\right|^{2}.\label{eq:Threshold}
\end{equation}

Let $\varphi=hx\tilde{w}$, and $\varphi_{r}$ and $\varphi_{i}$
represent the real and imaginary parts, respectively, of $\varphi$.
It can be readily seen that $\varphi$ is also a ratio of complex
Gaussian variables. Thus, the error probability can be expressed as
\begin{equation}
P_{b}=\stackrel[-\infty]{-\left|h\right|^{2}}{\int}\stackrel[-\infty]{\infty}{\int}f\left(\varphi\right)\mathrm{d}\varphi_{i}\mathrm{d}\varphi_{r},
\end{equation}
where
\begin{equation}
f\left(\varphi\right)=\tau\left|h\right|^{2}\left(\varphi_{r}^{2}+\varphi_{i}^{2}+\pi\tau\left|h\right|^{2}\right)^{-2}.\label{eq:f(phi)}
\end{equation}

To obtain the closed-form expression of $P_{b}$, the indefinite integral
of $f\left(\varphi\right)$ should be found. In \cite{Baxley2010_ratioCN_zeromean},
it is pointed out that the integral of the PDF \eqref{eq:f(phi)}
can be expressed in closed-form as
\begin{equation}
\iint f\left(\varphi\right)\mathrm{d}\varphi_{i}\mathrm{d}\varphi_{r}=\zeta\left(\varphi_{r},\varphi_{i}\right)+\zeta\left(\varphi_{i},\varphi_{r}\right)=G\left(\varphi_{r},\varphi_{i}\right),\label{eq:CDF}
\end{equation}
where
\begin{equation}
\zeta\left(\varphi_{r},\varphi_{i}\right)=\frac{\gamma\left(z_{i}\right)}{2\pi}\cdot\arctan\left(\frac{\varphi_{r}}{\sqrt{\pi\tau\left|h\right|^{2}+\varphi_{i}^{2}}}\right),
\end{equation}
and
\begin{equation}
\gamma\left(z_{i}\right)=\frac{\varphi_{i}}{\sqrt{\pi\tau\left|h\right|^{2}+\varphi_{i}^{2}}}.
\end{equation}

The BER $P_{b}$ is therefore expressed as
\begin{align}
P_{b} & =G\left(-\left|h\right|^{2},\infty\right)+G\left(-\infty,-\infty\right)-G\left(\infty,-\infty\right)-G\left(-\left|h\right|^{2},-\infty\right),\label{eq:Pb_G}
\end{align}
where the function $G$ is given by \eqref{eq:CDF}.

Substituting \eqref{eq:CDF} in \eqref{eq:Pb_G}, we obtain a \textit{closed-form
BER expression} as
\begin{align}
P_{b} & =\frac{1}{2}-\frac{1}{2\sqrt{\frac{\pi\tau}{\left|h\right|^{2}}+1}}.\label{eq:BER}
\end{align}
It can be seen that, by using the approximate linear channel model,
an exact closed-form BER expression can be found, which again shows
the benefit of channel linearization. Since we leverage the phase
information, it can be expected that \eqref{eq:BER} has lower BER
than the original ratio detector. The BERs of the proposed ratio detector
and the original ratio detector are numerically compared in Section
VI.

\section{Averaging and Coding}

While the previous results can be utilized directly to achieve AmBC
with high data rates, the strong interference from the direct link
limits performance. To overcome this challenge, three approaches are
proposed: 1) symbol averaging, 2) coding with hard and soft detection,
and 3) coding and interleaving. The formulations and advantages of
each of these are detailed next.

\subsection{Averaging}

Due to the development of the linear channel model \eqref{eq:linear channel model},
it is possible to implement straightforward averaging to enhance the
performance. While not an optimum approach, it is computationally
straightforward.

Expanding the duration of the Tag data is equivalent to transmit bit
repetition. The expansion duration, averaging period, or repetition
length is taken as $M$ symbols. It is assumed the channel coherence
time is $K$ symbols. To leverage the coherence of the channel, $M$
is less than $K$. For formulation convenience, however, we take $M=K$,
without loss of generality.

We assume that an arbitrary incoming bit stream at the tag is written
as $x_{1},x_{2},..$.. Since the bits are independent, we consider
a single bit $x_{k}$. The duration of bit $x_{k}$ is expanded $M$
times to form a transmit vector $\mathbf{x}_{k}=\left[x_{k};x_{k};...;x_{k}\right]$
of dimension $M$. At the Reader side, for $k$th received expanded
or repeated signal block, we have
\begin{equation}
\mathbf{y}_{k}=h_{k}\mathbf{x}_{k}+\tilde{\mathbf{w}}_{k},\label{eq:received_averaging}
\end{equation}
where $\mathbf{y}_{k}=\left[y_{1,k};y_{2,k};...;y_{m,K};...;y_{M,k}\right]$
represents the $k$th received block and $y_{m,k}$ represents the
$m$th received signal of the $k$th block, $h_{k}$ denotes the channel
realization of the $k$th block, and $\tilde{\mathbf{w}}_{k}=\left[\tilde{w}_{1,k};\tilde{w}_{2,k};...;\tilde{w}_{m,K};...;\tilde{w}_{M,K}\right]$
is the noise vector, where $\tilde{w}_{m,k}$ represents the noise
in $y_{m,k}$.

Due to the linearity and channel coherence, we can average over $y_{k}$
to decode $x_{k}$ to obtain
\begin{equation}
\hat{x}_{k}=\underset{x_{k}\in\left\{ -1,1\right\} }{\arg\max}\:f\left(\bar{y}_{k}\left|x_{k}\right.\right),\label{eq:averaging _ml}
\end{equation}
where $\bar{y}_{k}$ is the averaged received signal
\begin{equation}
\bar{y}_{k}=\frac{\stackrel[m=1]{M}{\sum}y_{m,k}}{M}.\label{eq:y_bar}
\end{equation}
Since the averaged signal $\bar{y}_{k}$ has the same distribution
as $y_{m,k},\forall m$, \eqref{eq:averaging _ml} can be written
as
\begin{equation}
\hat{x}_{k}=\underset{x_{k}\in\left\{ -1,1\right\} }{\arg\min}\:\left|\bar{y}_{k}-h_{k}x_{k}\right|^{2}.\label{eq:averaging}
\end{equation}

Since the noise is independent for each repeated bit, we can expect
the noise to be averaged while the desired signal adds coherently.
Therefore a doubling of $M$ will result in an approximately 3 dB
increase in the SNR (not exactly 3 dB because of the channel linearization
approximation) and the performance will improve.

\subsection{Coding without Interleaving}

Coding can be utilized as a more effective form of averaging. However,
to achieve the necessary gain in performance required, we need to
select very low coding rates such as $1/500$ or higher. Therefore,
conventional codes as used in terrestrial communications cannot be
utilized. For this reason, we utilize the conventional repetition
code with very low coding rates. Hard and soft detection are considered.

For the repetition code, the encoding and transmission processes are
the same as the averaging method. For the hard decision method, recall
that $x_{k}$ is related to $M$ received signals, and
\begin{equation}
y_{m,k}=h_{k}x_{k}+\tilde{w}_{m,k},\:m=1,...,M.
\end{equation}
For each signal, we can use the minimum distance detector \eqref{eq:minimum distance}
to decode it. Thus, for bit $x_{k}$, there will be $M$ decisions
and we take the majority as the final decision, which is
\begin{equation}
\hat{x}_{k}=\begin{cases}
1 & \textrm{if}\:\textrm{number of \ensuremath{1'}s}\geqslant-1's\\
-1 & \textrm{if}\:\textrm{number of \ensuremath{1'}s}<-1's
\end{cases}.\label{eq:hard decision}
\end{equation}

For soft decision detection, our detection rule is
\begin{align}
\hat{x}_{k} & =\underset{x_{k}\in\left\{ -1,1\right\} }{\arg\max}\:f\left(\mathbf{y}_{k}\left|x_{k}\right.\right)\nonumber \\
 & =\underset{x_{k}\in\left\{ -1,1\right\} }{\arg\max}\:\mathbf{\stackrel[\mathit{m=\mathrm{1}}]{\mathit{M}}{\prod}}f\left(y_{m,k}\left|x_{k}\right.\right)\nonumber \\
 & =\underset{x_{k}\in\left\{ -1,1\right\} }{\arg\min}\:\mathbf{\stackrel[\mathit{m=\mathrm{1}}]{\mathit{M}}{\sum}\log}\left(\left|y_{m,k}-h_{k}x_{k}\right|^{2}+\pi\tau_{k}\right),\label{eq:soft decision}
\end{align}
where $\tau_{k}=\left(1\left/\left|h_{1,k}^{\mathrm{SR}}\right|^{2}\right.+1\left/\left|h_{2,k}^{\mathrm{SR}}\right|^{2}\right.\right)N_{w}\left/\left(\pi P_{s}\right)\right.$,
and $h_{q,k}^{\mathrm{SR}}$ represents the small-scale fading between
the ambient source and the $q$th Reader antenna of the $k$th block.

Although the repetition code has the same operation as averaging at
the Tag side, since hard and soft decision detection methods consider
more information about the received signals, e.g., PDF for soft decision,
it can be expected that this coding scheme will outperform averaging.
The comparison is shown in Section VI.

\subsection{Coding with Interleaving}

To further combat deep fading, interleaving is combined with the repetition
code. During interleaving, the transmitted data is arranged over multiple
code blocks by the interleaver before backscattering. Due to this,
the block experiencing deep fading is spread out among multiple blocks.
When the Reader rearranges the blocks, the errors appear as independent
random errors or burst errors with short lengths, which are much easier
to detect.

Extending the formulations of Section IV.A and Section IV.B, we form
a block $\mathbf{X}$ as
\begin{equation}
\mathbf{X}=\left[\mathbf{x}_{1},\mathbf{x}_{2},...,\mathbf{x}_{k},...,\mathbf{x}_{K}\right].\label{eq:encoding_averaging-1}
\end{equation}
Applying interleaving, $\mathbf{X}$ is rearranged using a transpose
operation as
\begin{equation}
\mathbf{V}=\mathbf{X}^{T}=\left[\mathbf{v}_{1},\mathbf{v}_{2},...,\mathbf{v}_{m},...,\mathbf{v}_{M}\right],\label{eq:V}
\end{equation}
where $\mathbf{v}_{m}=\left[x_{1};x_{2};...;x_{k},...,x_{K}\right]$
is a $K$ dimensional vector representing the transmitting bit stream
during the $m$th block time. Now each repetition codeword is rearranged
into $M$ blocks.

\begin{algorithm}[t]
\begin{spacing}{1.1}
\textbf{Initialize:} $K$ is the channel coherent time and $M$ is
the codeword length of the repetition code. $K=M$. $x_{1},x_{2},...,x_{K}$
are modulated using BPSK.

1: Encoding: $\mathbf{X}=\left[\mathbf{x}_{1},\mathbf{x}_{2},...,\mathbf{x}_{k},...,\mathbf{x}_{K}\right]$,
$\mathbf{x}_{k}=\left[x_{k};x_{k};...;x_{k}\right]\in\mathbb{R}^{M}$

2:\textbf{ if} \textbf{without interleaving}

3:\textbf{ $\quad$for} $k=1:K$

4: \textbf{$\quad$$\quad$}Obtain $\mathbf{y}_{k}$ by \eqref{eq:received_averaging}

5: \textbf{$\quad$$\quad$if averaging}

6:\textbf{ $\quad$$\quad$$\quad$}Obtain $\bar{y}_{k}$ by \eqref{eq:y_bar}

7:\textbf{ $\quad$$\quad$$\quad$}Obtain $\hat{x}_{k}$ by \eqref{eq:averaging}

8: \textbf{$\quad$$\quad$end}

9: \textbf{$\quad$$\quad$if hard decision}

10:\textbf{$\quad$$\quad$$\quad$}Obtain $\hat{x}_{k}$ by \eqref{eq:hard decision}

11:\textbf{$\quad$$\;\,\;$end}

12:\textbf{$\quad$$\;\,\;$if soft decision}

13:\textbf{$\quad$$\quad$$\quad$}Obtain $\hat{x}_{k}$ by \eqref{eq:soft decision}

14:\textbf{$\quad$$\;\,\;$end}

15:\textbf{$\quad$end}

16:\textbf{ end}

17:\textbf{$\;\,$if} \textbf{with interleaving}

18:\textbf{$\;\,$}$\quad$Obtain $\mathbf{V}$, $\mathbf{Y}$, and
$\mathbf{R}$ by \eqref{eq:V}$-$\eqref{eq:R}

19:\textbf{$\;\,$}$\quad$\textbf{for} $k=1:K$

20:\textbf{$\;\,$}$\quad$\textbf{$\quad$}Repeat step $9-14$

21:\textbf{$\;\,$}$\quad$\textbf{end}

22:\textbf{$\;\,$end}

\textbf{Return} $\hat{\mathbf{x}}=\left[\hat{x}_{1};\hat{x}_{2};...;\hat{x}_{k},...,\hat{x}_{K}\right]$

\caption{Averaging, Coding and Interleaving}
\end{spacing}
\end{algorithm}

At the Reader side, after channel linearization, the received signal
matrix during $M$ coherent time slots is given by
\begin{equation}
\mathbf{Y}=\mathbf{V\mathrm{diag\left(\mathbf{h}\right)}}+\tilde{\mathbf{W}}=\left[\mathbf{y}_{1},\mathbf{y}_{2},...,\mathbf{y}_{m},...,\mathbf{y}_{M}\right],\label{eq:Y}
\end{equation}
where $\mathbf{y}_{m}=h_{m}\mathbf{v}_{m}+\tilde{\mathbf{w}}_{m}$
denotes the signal received in the $m$th block. $\mathbf{h}=\left[h_{1};h_{2};...,h_{m};...h_{M}\right]$
denotes the channel vector of $M$ blocks where $h_{m}$ denotes the
$m$th realization of channel $h$. $\tilde{\mathbf{W}}=\left[\tilde{\mathbf{w}}_{1},\tilde{\mathbf{w}}_{2},...,\tilde{\mathbf{w}}_{m},...\tilde{\mathbf{w}}_{M}\right]$
is the noise matrix where $\tilde{\mathbf{w}}_{m}=\left[\tilde{w}_{1,m};\tilde{w}_{2,m};...;\tilde{w}_{k,m};...;\tilde{w}_{K,m}\right]$
is the noise vector with $\tilde{w}_{k,m}$ representing the noise
of the $k$th received signal in the $m$th block. It can be seen
that $x_{k}$ experiences $M$ channels during interleaving, which
offers channel diversity to help detect the signal.

A deinterleaver is used to recover each repetition code. After rearranging
the received matrix $\mathbf{Y}$, we have
\begin{equation}
\mathbf{R}=\mathbf{Y}^{T}=\left[\mathbf{r}_{1},\mathbf{r}_{2},...,\mathbf{r}_{k},...,\mathbf{r}_{K}\right],\label{eq:R}
\end{equation}
where
\begin{equation}
\mathbf{r}_{k}=\left[r_{k,1};r_{k,2};...,r_{k,m};...r_{k,M}\right]=\mathbf{h}x_{k}+\mathbf{u}_{k},
\end{equation}
with $\mathbf{u}_{k}=\left[\tilde{w}_{k,1};\tilde{w}_{k,2};...;\tilde{w}_{k,m};...;\tilde{w}_{k,M}\right]$
representing the noise vector.

Since each repetition code is independent, we can use $\mathbf{r}_{1}$
to $\mathbf{r}_{K}$ to decode $x_{1}$ to $x_{K}$. Correspondingly
hard and soft decision detection can then be applied similarly to
that without interleaving.

Compared with Section IV.A and Section IV.B, it can be seen that since
interleaving is utilized, the encoding and detection complexity is
increased for this scheme. However, the channel diversity can significantly
lower BER, which will be verified in Section VI.

The overall algorithm for the transmission and decoding process for
the three proposed schemes is summarized in Algorithm 1. It should
be noted that the channel vector $\mathbf{h}$ is obtained based on
the proposed linear channel model, without which the soft decision
and hard decision processes will be difficult to implement.

\section{Optimum Ratio Selection Scheme}

In this section, we generalize the ratio detector to Readers with
$Q>2$ antennas. With $Q>2$ Reader antennas, there are a variety
of ratios that can be considered, as well as many methods to utilize
them. For example, the ratio need not be the straightforward ratio
of a pair. The resulting ratios can also be combined in various ways
including some form of maximal ratio combining. The full investigation
of detectors for dealing with $Q>2$ receive antennas can therefore
be seen to be a significant undertaking.

To demonstrate the potential of our proposed ratio detector, and averaging,
coding, and interleaving techniques with $Q>2$, we restrict our approach
to the most straightforward system and leave the investigation of
the complete generalization to future work. As such we restrict the
ratios to be between pairs of antenna branches and then propose an
optimal ratio selection scheme to leverage selection diversity. Even
this straightforward approach provides significant improvements in
performance as shown later in the simulation results section.

Our goal is to select the optimal ratio to minimize the BER \eqref{eq:BER}.
Accordingly, the optimization problem can be formulated as
\begin{align}
\mathbf{P1}:\underset{i,j}{\mathrm{min}}\;\; & \frac{1}{2}-\frac{1}{2\sqrt{\frac{\pi\tau_{i,j}}{\left|h_{i,j}\right|^{2}}+1}}\\
\mathrm{s.t.}\;\;\; & i,j\in\left\{ 1,2,...,Q\right\} ,i\neq j
\end{align}
where $\tau_{i,j}$ and $h_{i,j}$ represent the value of $\tau$
and $h$ calculated with the $i$th and $j$th antenna pairs, $i,j\in\left\{ 1,2,...,Q\right\} ,i\neq j$.

To solve this problem, we first observe that $P_{b}$ monotonically
increases with $\eta_{i,j}$, which is given by
\begin{equation}
\eta_{i,j}=\frac{\pi P_{s}\left|g\right|^{2}}{N_{w}}\cdot\frac{\tau_{i,j}}{\left|h_{i,j}\right|^{2}}=\frac{\frac{1}{\left|h_{i}^{\mathrm{SR}}\right|^{2}}+\frac{1}{\left|h_{j}^{\mathrm{SR}}\right|^{2}}}{\left|\frac{h_{i}^{\mathrm{TR}}}{h_{i}^{\mathrm{SR}}}-\frac{h_{j}^{\mathrm{TR}}}{h_{j}^{\mathrm{SR}}}\right|^{2}},\label{eq:yita_ij}
\end{equation}
where subscript $i,j$ denotes a parameter related to the $i,j$th
antenna pair. Since $\eta$ is only related to the channel parameters
of the two branches in the ratio, by selecting a ratio with the minimum
value of $\eta$ among all the ratios, a lower BER can be obtained.

With $Q$ Reader antennas, $Q\left(Q-1\right)$ ratio pairs can be
formed. It should be noted that for each ratio pair, swapping the
numerator and denominator does not affect the value of $\eta$. Thus
we further restrict to $i<j$, and only $Q\left(Q-1\right)\left/2\right.$
ratios need to be searched.

Subsequently, our antenna selection scheme can be described as follows.
First, we compute the value of $\eta$ according to \eqref{eq:h}
and \eqref{eq:tao} among $Q\left(Q-1\right)\left/2\right.$ ratios.
Next, we select the ratio $\lambda_{\mathrm{opt}}$ which has the
minimum value of $\eta$. The detailed optimal ratio selection scheme
is shown in Algorithm 2.

\begin{algorithm}[t]
\begin{spacing}{1.1}
\textbf{Initialize:} Channel parameters $h_{q}^{\mathrm{SR}}$, $h_{q}^{\mathrm{ST}}$,
$q\in\left\{ 1,2,...,Q\right\} $. $\mathrm{\eta}_{\mathrm{min}}$
is sufficiently large.

1: \textbf{for} $j=2:Q$

2:\textbf{ $\quad$for }$i=1:j-1$

3: \textbf{$\quad$$\quad$}Calculate $\eta_{i,j}$ using \eqref{eq:yita_ij}

4: \textbf{$\quad$$\quad$if }$\eta_{\mathrm{min}}\geqslant\eta_{i,j}$

5: \textbf{$\quad$$\quad$}$\quad$Update $\eta_{\mathrm{min}}$
by $\eta_{i,j}$

6: \textbf{$\quad$$\quad$end}

7: \textbf{$\quad$end}

8: \textbf{end}

9: Select antenna $i,j$ according to $\eta_{\mathrm{min}}$

10: Calculate the optimal ratio $\lambda_{\mathrm{opt}}=\bar{z}_{i}\left(n\right)\left/\bar{z}_{j}\left(n\right)\right.$

\textbf{Return} $\lambda_{\mathrm{opt}}$

\caption{Optimal Ratio Selection Scheme}
\end{spacing}
\end{algorithm}

\section{Simulation Results}

In the simulation results presented, we assume that all the small-scale
channel fading coefficients $h_{q}^{\mathrm{SR}}$, $h_{q}^{\mathrm{TR}}$
and $h^{\mathrm{ST}}$ follow a $\mathcal{CN}\left(0,1\right)$ distribution.
The hardware implementation loss by the Tag, $\alpha$, is set as
$1.1$ dB \cite{passiveWIFI_Kellogg_2016} and the relative SNR $\Delta\gamma=40$
dB \cite{Wenjing_FIRST_2021}. The Monte Carlo method is used to find
the BER.

\subsection{Comparison with the Original Ratio Detector}

\begin{figure}[t]
\begin{centering}
\includegraphics[scale=0.48]{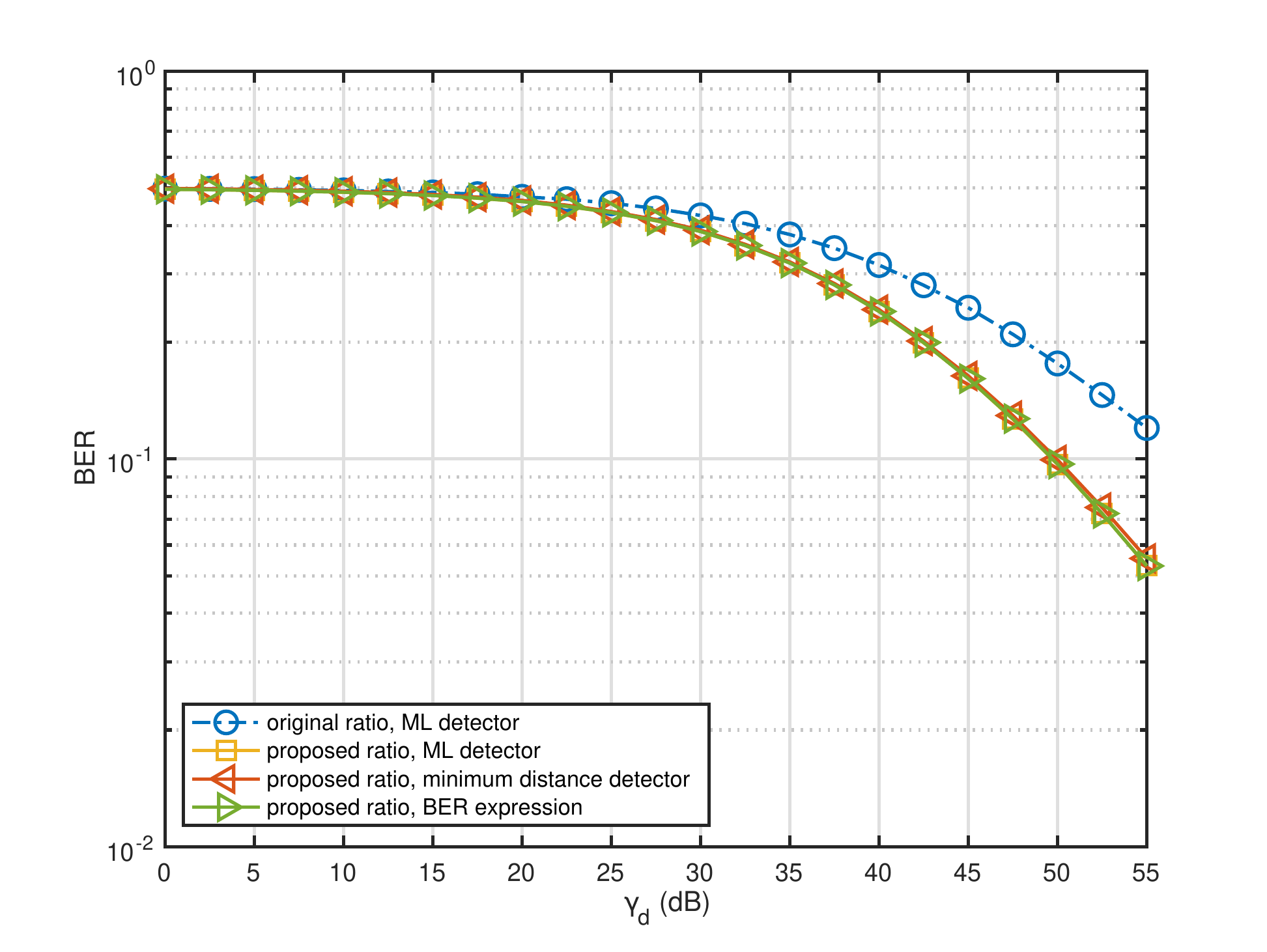}
\par\end{centering}
\centering{}\caption{\label{fig:umo vs our}BER versus the direct link SNR of the original
and proposed ratio detectors.}
\end{figure}

We first compare the detection performance of the original ratio detector
\cite{expand_umo_Ma_2018} and our proposed ratio detector. The simulated
BERs versus direct link SNR of the two detectors are shown in Fig.
\ref{fig:umo vs our}. From Fig. \ref{fig:umo vs our}, three observations
can be made.

\textit{First,} for the proposed ratio detector, it can be observed
that the optimal ML detector \eqref{eq:ML_before_linearize} based
on \eqref{eq:advanced_channel_model} and the minimum distance detector
\eqref{eq:minimum distance} based on the linearized channel \eqref{eq:linear channel model}
have almost the same BERs. This shows that the proposed linearized
channel model \eqref{eq:linear channel model} well approximates the
channel \eqref{eq:advanced_channel_model}. In addition, the BER performance
of the proposed ratio detector matches well with the theoretical analysis
result \eqref{eq:BER}, which validates the correctness of our BER
analysis.

\textit{Second}, comparing the BERs of the original ratio detector
and our proposed detector, we can find that at the same SNR point,
the proposed ratio detector has a lower BER. This is because preserving
phase information gives the detector more information.

\textit{Third}, it can be seen in Fig. \ref{fig:umo vs our} that
for both detectors, the BER curves decrease significantly when the
SNR is higher than 25 dB. This is because when the data rate for the
ratio detector is as high as the rate of the ambient signals, at each
Tag (or ambient) symbol period, the average power of the backscattered
signal is low, which results in a low receive SNR. Only when the direct
link SNR is high enough, it can be ensured that the receive SNR is
as high as the level that can detect the bit error. This can be overcome
by averaging or coding.

It is worth noted that for our proposed ratio detector which leverages
the linear channel model, the ML detector is simplified to be the
minimum distance detector, which greatly reduces the detection complexity.
This also allows a closed-form BER expression to be obtained. Furthermore,
benefiting from the channel linearization, averaging and coding schemes
can be utilized to further enhance the system performance, as will
be shown in the next subsection.

\subsection{Evaluation of Averaging and Coding}

\begin{figure}[t]
\centering{}\includegraphics[scale=0.48]{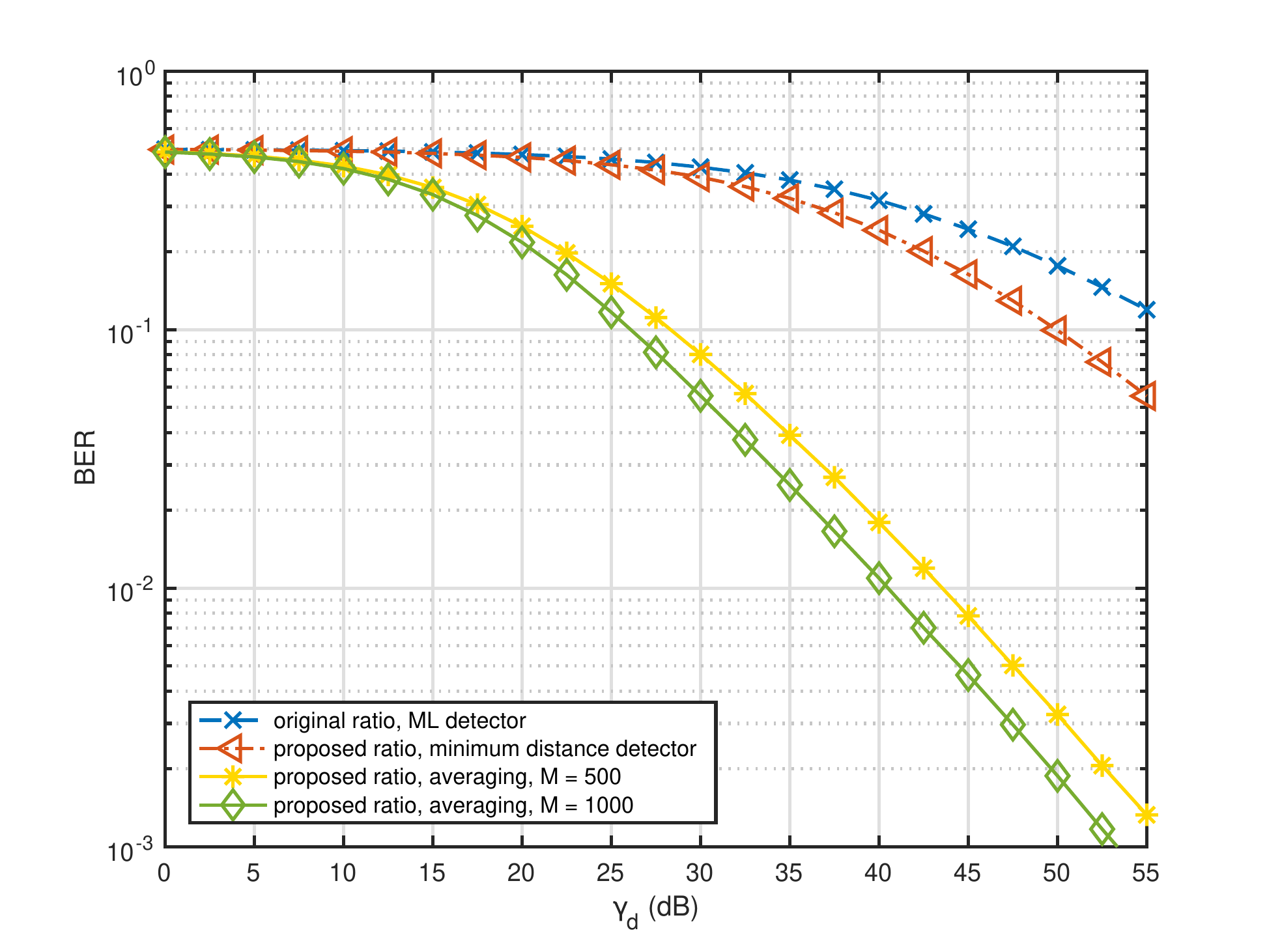}\caption{\label{fig:averaging}BER versus the direct link SNR of the original
and proposed ratio detectors with (for $M=$ 500 and 1000) and without
averaging.}
\end{figure}

Herein we evaluate the BER performance of the averaging and coding
scheme proposed in Section IV based on the accurate approximate linear
channel model \eqref{eq:linear channel model}.

In Fig. \ref{fig:averaging}, the averaging method proposed in Section
IV.A is evaluated for the proposed ratio detector, where the symbol
duration is $M$ times longer than the ambient signal period. The
performance of the original ratio ML detector as well as the proposed
ratio detector using minimum distance detection with no averaging
is also shown as a reference. It can be seen that after averaging,
the detection performance has been greatly improved. Comparing the
curves of $M=500$ and $M=1000$, it can be seen that doubling the
codeword length left shifts the BER approximately by 3 dB. However,
the BER is still not ideal at the low SNR region. This is because
deep fading may cause all the received samples to be lost during the
bad channel condition.

\begin{figure}[t]
\centering{}\includegraphics[scale=0.48]{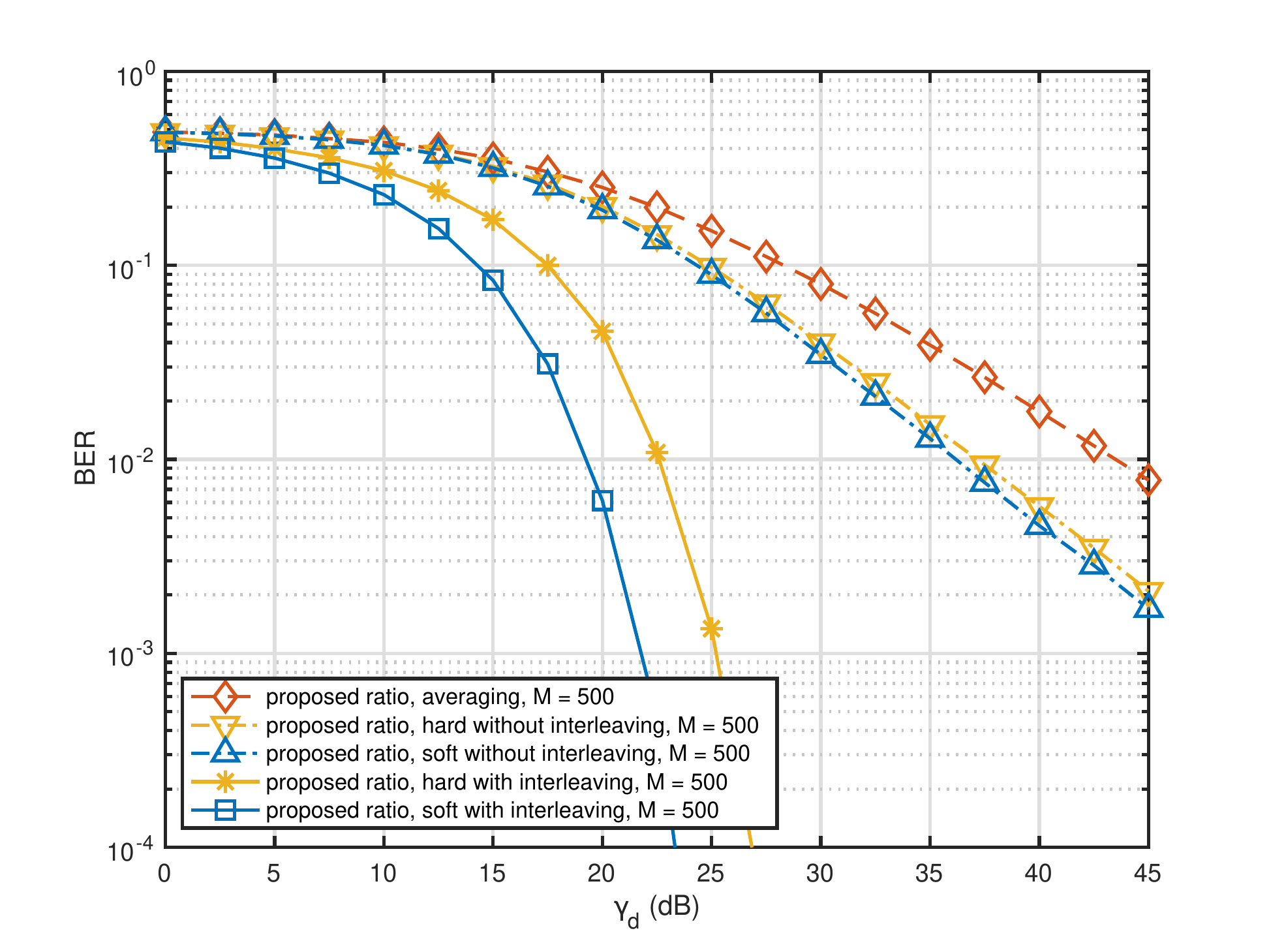}\caption{\label{fig:different methods}BER versus the direct link SNR of the
proposed ratio detector with and without interleaving when the averaging
period (or length of repetition code) $M=500$.}
\end{figure}

In Fig. \ref{fig:different methods}, hard and soft decoding with
and without interleaving results are provided. The BER performance
of averaging is also shown as a reference. The length of averaging
duration or length of the repetition code is $M=500$ for all the
methods. In Fig. \ref{fig:different methods}, three observations
can be made.

\textit{First}, it can be seen that although both repetition code
and averaging have the same operation at the Tag side, which is to
expand the duration of data (or repetition), repetition code has lower
BER than averaging. This indicates that both the hard and soft decision
methods have better detection performance compared with averaging
the samples out.

\textit{Second}, since the soft decision method considers the joint
PDF of the received code word, it can be observed that it outperforms
the hard decision decoding. Nevertheless, although the BER of repetition
code is lower than averaging, deep fading still has considerable negative
effect on detection so that the performance of using repetition code
only is still not desirable.

\textit{Third}, after combining the repetition code with interleaving,
it can be observed that the performance improvement is impressive.
The BER curves of both the soft decision method and the hard decision
method drop sharply and can achieve BER at $10^{-4}$ level around
24 dB and 26 dB, respectively, which shows the efficiency of the encoding
and detection method proposed in Section IV.C.

\subsection{Comparison with Averaging-based Energy Detector}

\begin{figure}[t]
\begin{centering}
\includegraphics[scale=0.48]{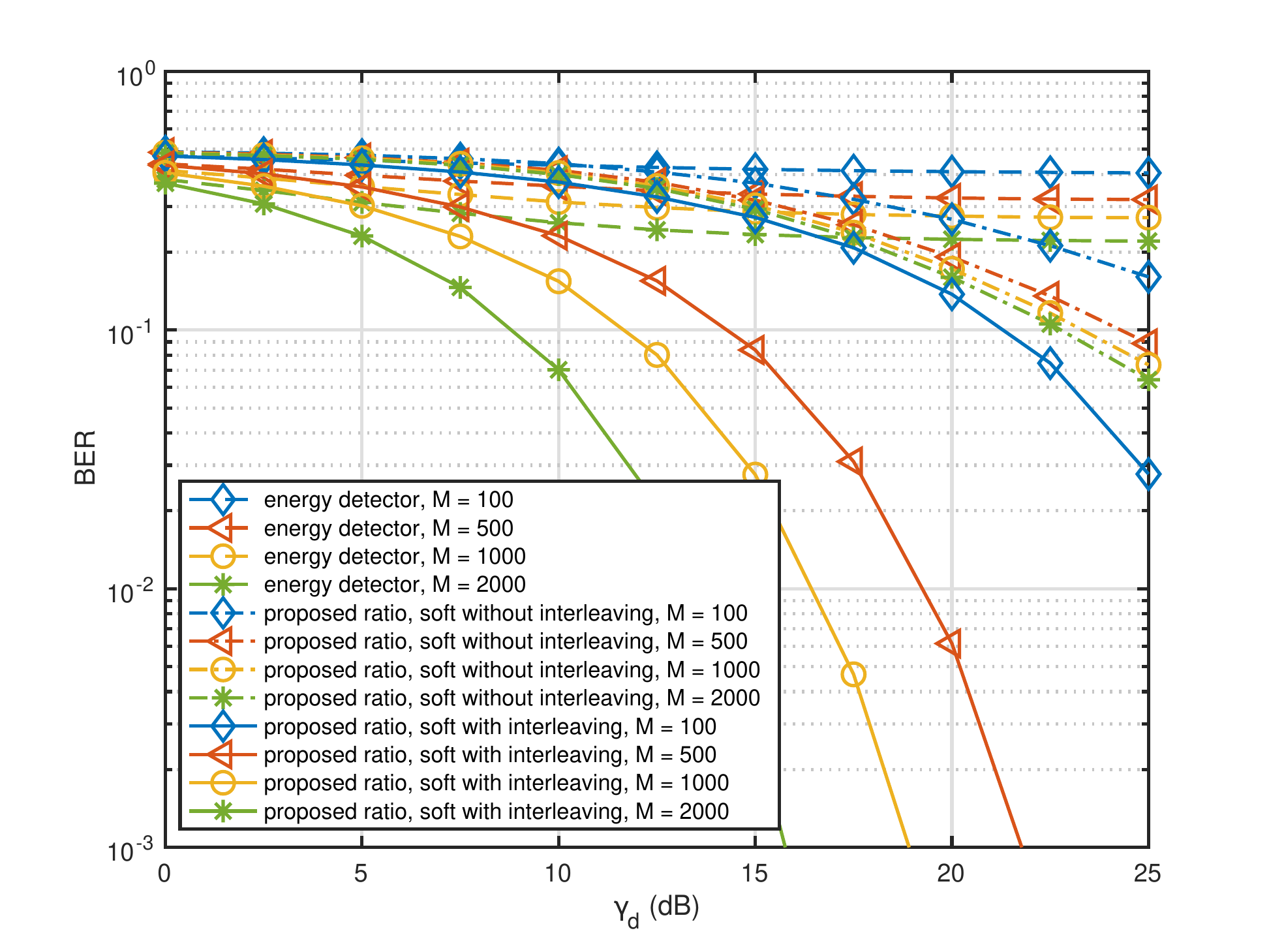}
\par\end{centering}
\caption{\label{fig:repetition code_vs_energy}BER versus direct link SNR of
averaging-based energy detector and the proposed ratio detector in
Section IV; averaging period (or length of repetition code) $M=100,500,1000,2000$.}
\end{figure}

Fig. \ref{fig:repetition code_vs_energy} compares the performance
of our proposed ratio detector and that of the averaging-based energy
detection from our previous work \cite{Wenjing_FIRST_2021}. For our
proposed ratio detector, we use the repetition code to enhance its
performance and adopt soft decision method to detect the data. The
scenarios with and without interleaving are both considered. The averaging
period (or length of repetition code) $M=100,500,1000,2000$. From
Fig. \ref{fig:repetition code_vs_energy} we have the following observations.

\textit{First}, for both kinds of detectors, the BERs decrease with
increase in $M$ when the direct link SNR is fixed. However, the ratio
detector is more responsive to change in $M$. When $M$ is fixed,
the BER for averaging-based energy detection will not decrease when
the direct link SNR is large, in other words, there is an error floor.
But for the proposed ratio detector, the error floor disappears; on
the contrary, the slope of the BER curve increases as the direct link
SNR increases.

\textit{Second}, comparing the performance of the averaging-based
energy detector and ratio detector having the same $M$, we can find
that without interleaving, and with sufficient SNR ($>20$ dB), the
proposed ratio detector achieves a lower BER than the averaging-based
energy detector. However, after interleaving, our proposed detector
always outperforms the averaging-based energy detector and the performance
gap is impressive. It can be seen that when $\gamma_{d}=15$ dB and
$M=1000$, the BER of the proposed detector is almost 100 times lower
than energy detectors. Besides, the performance gap increases with
increasing direct link SNR, which shows the superiority of our proposed
ratio detector and scheme developed in Section IV.C.

\begin{figure}[t]
\begin{centering}
\includegraphics[scale=0.48]{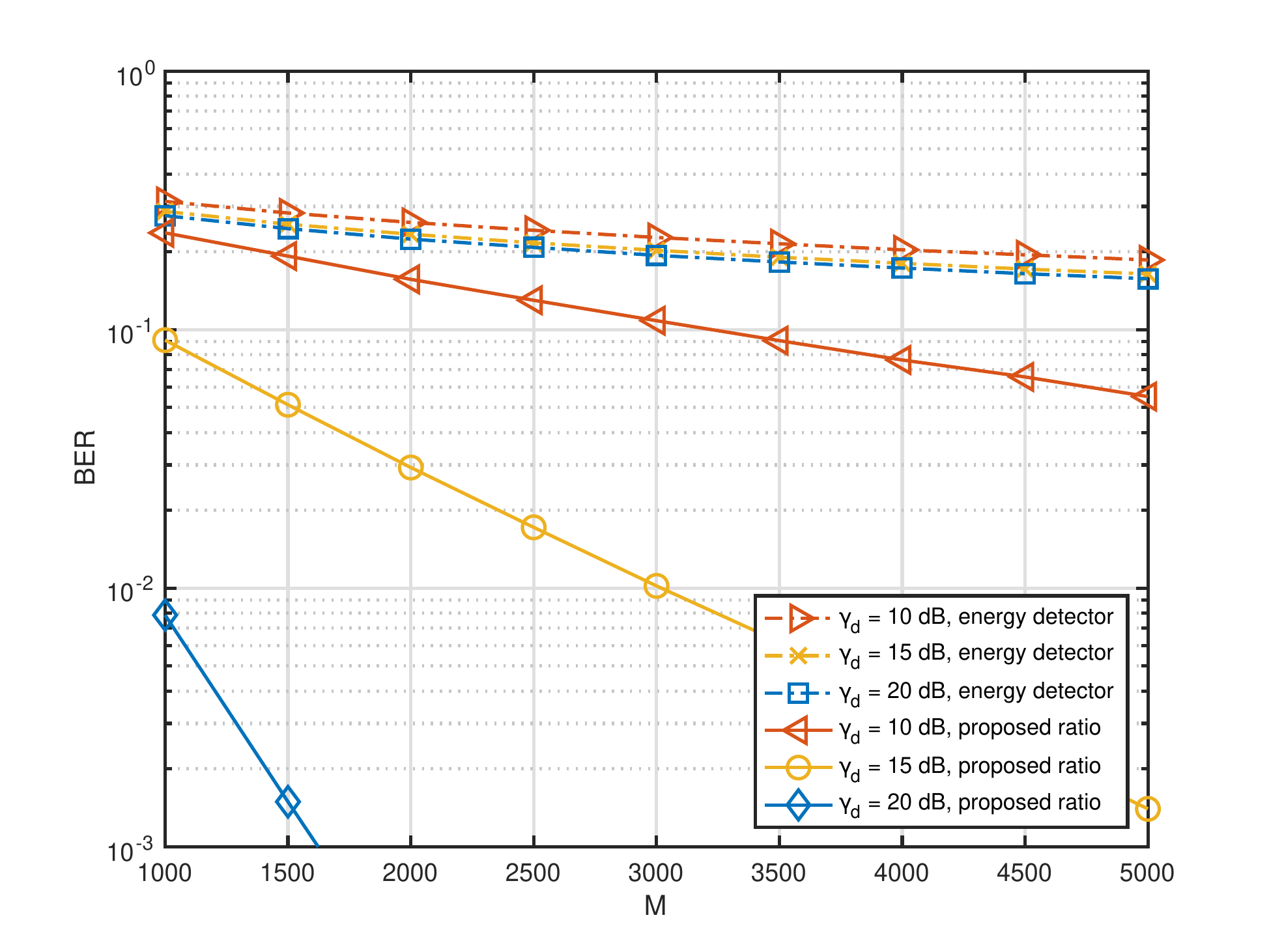}
\par\end{centering}
\caption{\label{fig:N_vs_BER}BER versus $M$ of averaging-based energy detector
and our proposed ratio detector with repetition code; direct link
SNR $=10,15,20$ dB.}
\end{figure}

One of the most important advantages of the ratio detector is the
increased data rate. Therefore, in Fig. \ref{fig:N_vs_BER}, we compare
the BER versus $M$ of the averaging-based energy detector and our
proposed detector with repetition code. For our proposed ratio detector,
hard decision decoding with interleaving have been adopted. For both
of the detectors, the single-antenna Tag and dual-antenna Reader is
utilized. We take three different direct link SNRs: $10,15,20$ dB,
and the following two observations can be made from Fig. \ref{fig:N_vs_BER}.

\textit{First}, when $M$ is fixed, both of the detectors achieve
lower BER when SNR is increased. However, the proposed ratio detector
is more responsive to the change in direct link SNR. It can be observed
that for the ratio detector, the performance gap between each curve
having a different SNR is much larger than that of averaging-based
energy detection.

\textit{Second}, for both the detectors, increasing $M$ can improve
system performance when the direct link SNR is fixed. Nevertheless,
with the same direct link SNR, the proposed ratio detector has a much
lower BER compared with the averaging-based energy detector and the
performance gap becomes larger with increases in $M$.

\subsection{Ratio Selection Scheme and Phase Compensation}

\begin{figure}[t]
\begin{centering}
\includegraphics[scale=0.48]{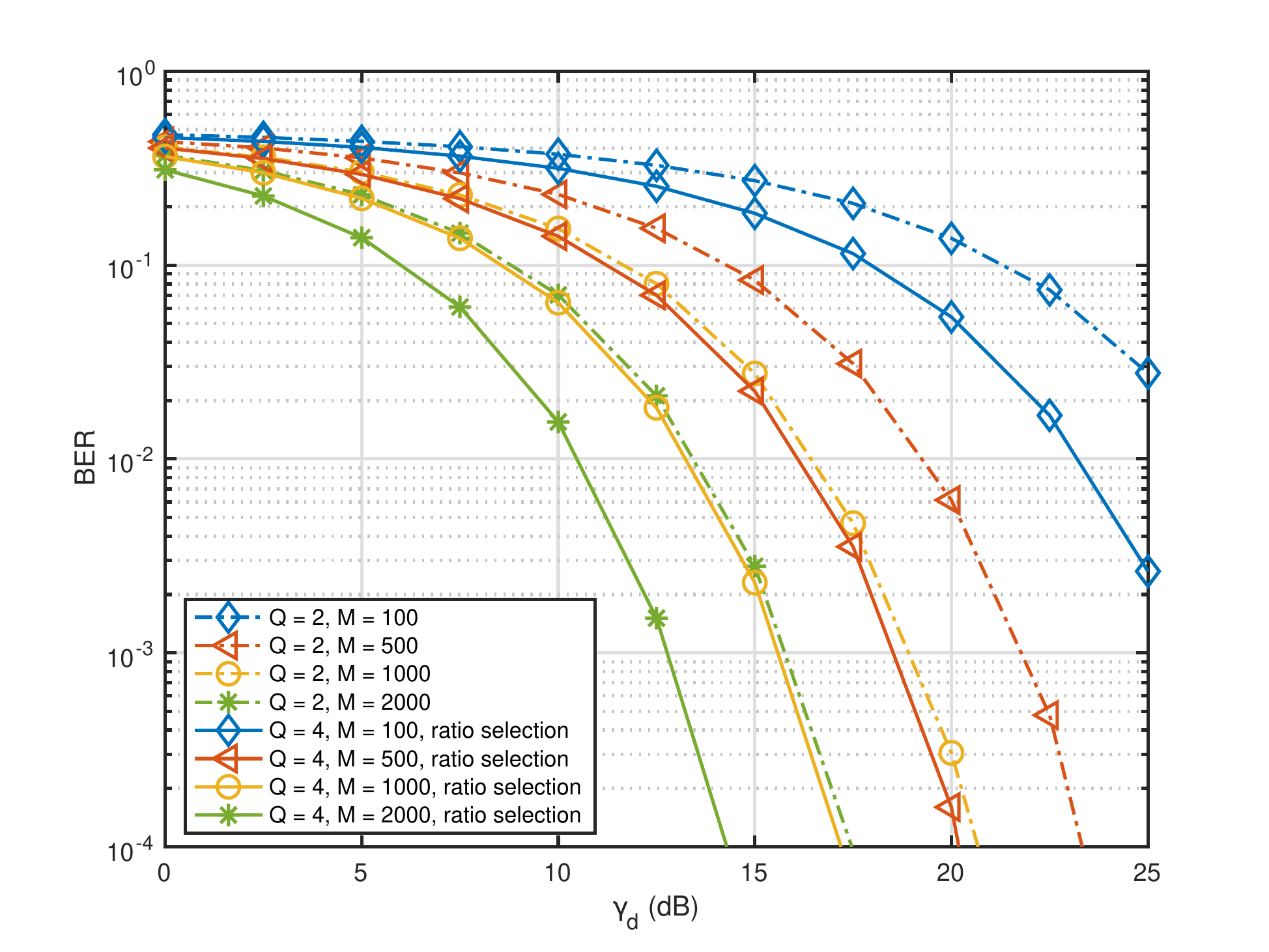}
\par\end{centering}
\caption{\label{fig:ratio selection}BER versus direct link SNR of four-antenna
Reader based on optimum ratio selection scheme; $M=100,500,1000,2000$.}
\end{figure}

To provide quantification of the gains that can be achieved with more
than two antennas at the Reader, the BER performance of a 4-antenna
($Q=4)$ Reader utilizing ratio selection is shown in Fig. \ref{fig:ratio selection}.
The BER curve of a $Q=2$ antenna system is shown as a reference.
Both use interleaving to combat deep-fading and use the soft decision
method to detect the Tag data. From Fig. \ref{fig:ratio selection}
it can be seen that by implementing multiple antennas on the Reader,
the ratio that has minimum $\eta$ can be selected and the BER is
greatly lowered compared with BER of the $Q=2$ Reader. For $M=500$
and $M=1000$ cases, it can be observed that the $Q=4$ Reader using
ratio selection scheme has better performance than $Q=2$ Reader\textquoteright s
$M=1000$ and $M=2000$ cases, correspondingly. This validates the
effectiveness of our proposed ratio selection scheme.

\begin{figure}[t]
\begin{centering}
\includegraphics[scale=0.48]{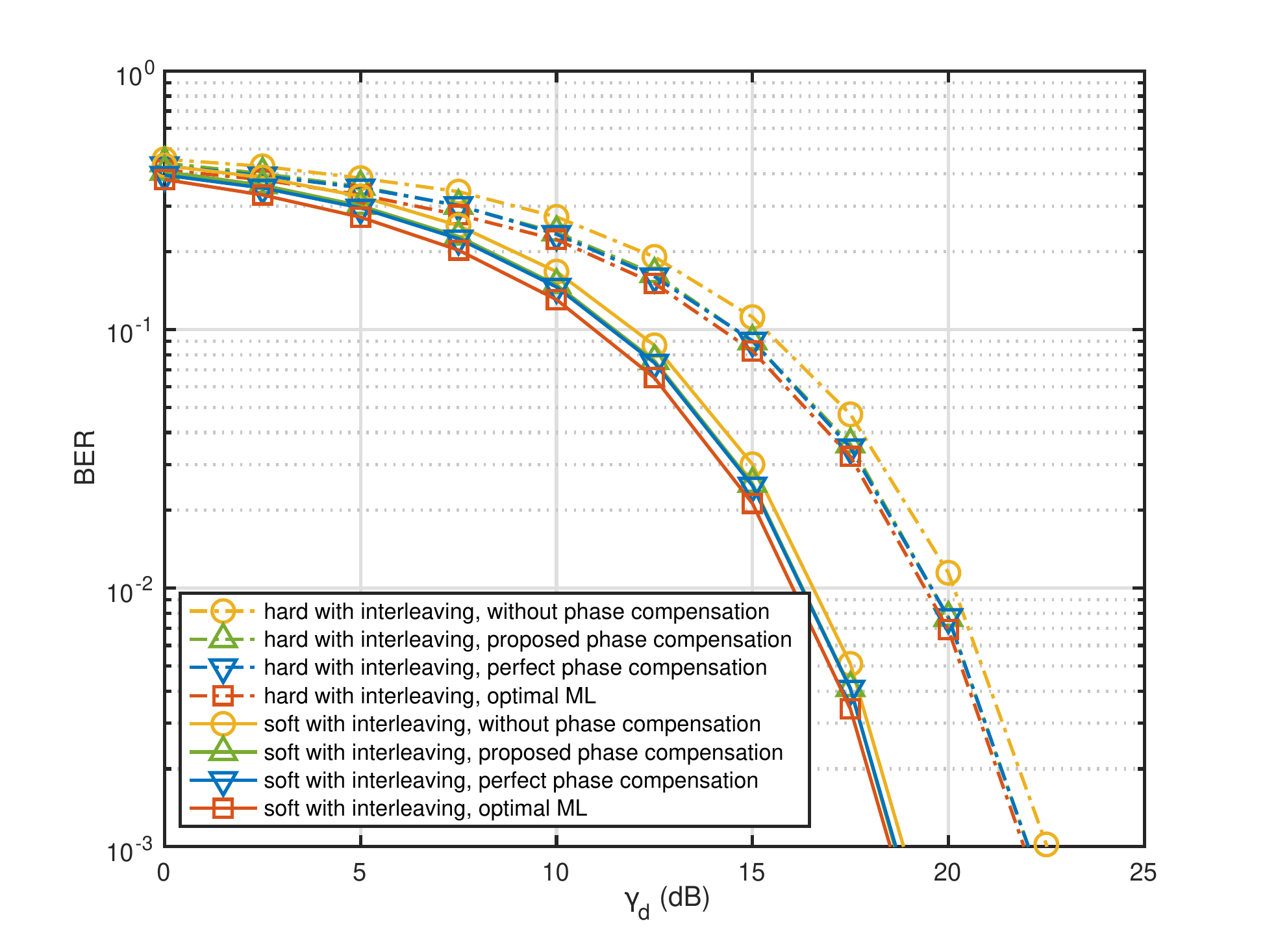}
\par\end{centering}
\caption{\label{fig:phase}Effect of phase compensation. BERs versus direct
link SNR of our proposed detector with repetition code under different
cases, $M=1000.$}
\end{figure}

Finally, for completeness, we also investigate the effectiveness of
phase compensation and the results are shown in Fig. \ref{fig:phase}.
In the results, both hard and soft decision decoding with interleaving
have been adopted and the repetition codeword length is $M=1000.$
As discussed, the left-hand side of \eqref{eq:linearized model_without_phase}
is not always equal to its right-hand side because a $2\pi$ phase
shift may occur. In order to get the accurate linear channel model,
it is essential to rectify the phase shift. In Fig. \ref{fig:phase},
we detect the Tag data under three different cases and compare them
with the performance of optimal ML detector \eqref{eq:ML_before_linearize}.

It can be seen that for both soft and hard decision cases, the BER
performance is worst if we detect the Tag data without any phase compensation.
This is because a $2\pi$ phase shift may occur and degrade performance.
After implementing the phase compensation scheme proposed in \eqref{eq:phase shift},
we can obtain almost the same performance as perfect phase compensation,
which implies that the accuracy of the proposed phase compensation
is very high. Comparing the BER curves of our proposed phase compensation
scheme and perfect phase compensation with the optimal performance,
it can be seen that there is a small gap, which results from the approximation
error of linearization. In addition, it can be seen that for both
soft and hard decision cases, this performance gap decreases with
the increase of SNR, since higher SNR provides a more accurate channel
linearization.

Overall, the proposed phase compensation scheme can effectively solve
the phase issue caused by taking the principal value of the logarithm
and thus enhance the detection performance of the proposed ratio detector.

\section{Conclusion}

In this paper, we have proposed an AmBC system with an efficient ratio
detector to overcome the drawbacks of conventional averaging-based
energy detectors. Unlike original ratio detectors that use the magnitude
ratio of the signals between two Reader antennas, we have utilized
the complex ratio so that phase information can be preserved. This
allows us to devise an accurate linear model for the ratio detector,
which can be utilized to open up the use of more standard detection
techniques in AmBC systems. Based on the constructed linear channel
model, we have proposed the minimum distance detector and derived
closed-form expressions for the BER. Averaging, coding, and interleaving
can also be straightforwardly applied. Besides, the results are general,
allowing any number of Reader antennas to be utilized in the approach.
Simulation results demonstrate that the proposed detector is better
than averaging-based energy detectors and original ratio detectors.
These results indicate that the proposed technique is potentially
useful for future AmBC systems.


\end{document}